\documentclass[12pt]{article}

\usepackage{amssymb}

\def\be{\begin{equation}}
\def\ee{\end{equation}}
\def\bea{\begin{eqnarray}}
\def\eea{\end{eqnarray}}

\def\d{\partial}
\def\D{\nabla}

\def\t{\tilde}

\def\g{\gamma}

\def\icv{\nu_1,...,\nu_m }
\def\ict{\mu_1,...,\mu_l}

\def\ic{^{\rho}_{\mu\nu}}

\def\cM{{\cal M}}

\def\g{\gamma}

\def\D{\nabla}

\begin{document}
\begin{titlepage}

\begin{flushright}
IFT-UAM/CSIC-97-01\\
hep-th/9711199 \\
\end{flushright}

\vspace{3cm}
\begin{center}
\large{{\bf KILLINGS, DUALITY AND CHARACTERISTIC POLYNOMIALS } } \\
\vspace{1.0cm}

\large{\bf {Enrique \'Alvarez, Javier Borlaf and Jos\'e H. Le\'on}} \\
\vspace{5mm}

Instituto de Fisica Te\'orica, CXVI \\
and\\
Departamento de Fisica Te\'orica,CXI\\
Universidad Aut\'onoma, 28049 Madrid, Spain \\
enrial@daniel.ft.uam.es,javier@delta.ft.uam.es,
jhleon@delta.ft.uam.es
 
\end{center}
\vspace{1.5cm}

\begin{abstract}

In this paper the complete geometrical setting of (lowest order) abelian
T-duality is explored with the help of some new 
geometrical tools (the {\it
reduced formalism}). In particular, all invariant polynomials 
(the {\it integrands} of the characteristic classes) can be explicitly 
computed for the dual model in terms of quantities pertaining to the
original one and with the help of the 
{\it canonical connection} whose intrinsic
characterization is given.
Using our formalism the physically, and T-duality {\it invariant},
relevant result that top 
forms are zero when there is an isometry without fixed 
points is esily proved.

\end{abstract}
\vspace{15mm}

\noindent

\end{titlepage}
\setcounter{footnote}{0}

\section{Introduction}

T-duality (cf.\cite{gpr}\cite{aagl}) has become a standard tool in the
general 
exploration
of the space of vacua in String Theory (apparently best thought of as
coming 
from an 11-dimensional M-theory ), albeit almost
always in the somewhat trivial setting of toroidal compactifications.
\par
Although Buscher's formulas are known not to be exact except in the
simplest cases
(i.e., generically there are $\alpha ' $ corrections to them), it is
nevertheless very 
interesting to get as much information as possible on the ``classical''
geometry
of the dual target space.
\par
In the present paper we generalize (to the case in which there is a
n-dimensional
abelian group of isometries) and develop further a formalism first
discussed
by one of us in \cite{b}, thus allowing to compute all interesting 
geometrical quantities
of the dual space. Ariadne's thread will consist in
exploiting the residual
gauge invariance in adapted coordinates.
\par
As a subproduct all invariant polynomials in the dual space are explicitly 
determined, yielding some general conclusions on the vanishing of certain 
topological invariants when they are zero. Some new T-invariants (i.e.,
scalars
under T-duality) also stem from the analysis.
\par
Our results are strongest when all points have trivial isotropy group;
that is, when the isometry does not have any fixed points, which in the
euclidean
signature is the same thing as to say that the Killing vector never has
zero modulus.

It is a well known fact than when performing a T-duality transformation,
the geometry
of the dual space (insofar as this concept makes sense) can be wildly
different
from the one of the original space. There are two qualifications. First,
Buscher's formulas
are expected to receive corrections in all but the simplest cases
\cite{abl}. Second,
the probes to be used to explore the dual geometry are not neccessarily the
same
as the ones to be used in the original one (cf. \cite{aagbl} for some
 comments on the
operator mapping).\\

It is nevertheless of great interest to determine the ``classical''
geometry of the 
dual space in as precise a manner as possible. (cf. \cite{orlando} for
previous attempts
in this direction).
\\
\section{The Reduced Formalism}

In this section we will give a very brief summary about the {\it
Reduced
Formalism} introduced in \cite{b} .
We shall assume that the Target Space manifold M is invariant under a
n-dimensional
 abelian group of isometries, and the corresponding Killing vectors will be
denoted by ${k^{\mu}_{a}}$ where
${\mu,\nu = 0,1,...,D-1}$, ${a,b = 0,...,n-1}$ and D is the dimension of M.
We start with two conditions.The first one, we have a 
set  $\Sigma$ of G-invariant
tensors, 
$V$ ,  characterized by:
\medskip
\be
\label{c1}
{\cal L}_{k_a} V = 0
\ee
which in adapted coordinates ${x^{i},x^{a}}$ ($i = n,...,D-1$) reduces to  
$\d_{a}V = 0$.\\

The second condition consist on having a connection compatible with
the first one, which implies

\be
\label{lie2}
({\cal L}_{k_a}\D)_{\mu\nu}^{\sigma}=
[{\cal L}_{k_a} \, , \D_{\mu}] _{\nu}^{\sigma}=
K^{\alpha}_{a}R_{\alpha\mu\nu}^{\sigma} \, +
\D_{\mu}\D_{\nu}K^{\sigma}_{a}
\, + 2\D_{\mu}(K^{\alpha}_{a}T_{\alpha\,\nu}^{\sigma})\equiv 0
\ee
\\
$R^{\sigma}_{\alpha\mu\nu}$ is the curvature of the connection
$\Gamma_{\lambda\beta}^{\delta}$ and $T\ic$ is the corresponding torsion.
(\ref{lie2}) is nothing but $\partial_{a}\Gamma_{\mu\nu}^{\sigma}=0$ 
in adapted coordinates.
\par
Given a set of commuting vector fields (in our case, the Killings), 
$k_{a}^{\mu}$, there always exists 
a system of coordinates ({\it adapted coordinates}) such that
$ k^{\mu}_{a} = \delta_{a}^{\mu}$ , i. e., $ k_{a}\equiv
{\partial\over\partial x^a}$.
(cf.\cite{lichne}).
These conditions do not determine completely the system of coordinates;
the residual gauge group actually consists in arbitrary compositions
of transverse diffeomorphisms ($x^i\, ' = f^i (x^j)$) with redefinitions of
the 
ignorable coordinates themselves:
\bea
\label{ad}
&x^{'i} = x^i \nonumber\\  
&x^{'a} = x^{a} + \Lambda^a (x^j)
\eea

\hfill

Tensors in  $\Sigma$ transform linearly under (\ref{ad}):

\be
\label{tt}
V' = J(\d \Lambda)V
\ee

\hfill

If we had at our disposal some {\it transverse gauge fields} 
$A^{a}_i (x^j) $
, i.e., fields transforming under
(\ref{ad}) as: 

\be
\label{gauge}
A'^{a}_{i}(x^j)=A^{a}_{i}(x^j) - \d_{i}\Lambda^{a}(x^j)
\ee

\hfill

Then we could associate the {\it reduced tensor} $v$ to each
tensor $V$ by:
\be
\label{rt}
v \equiv J(A)V
\ee

\hfill

Reduced tensors are invariant under (\ref{ad}), i.e.,
$v'= J(A-\d \Lambda)J(\d\Lambda)V = v$, because $J$ 
provides a representation of the abelian group G in the 
space of tensors characterized by (\ref{c1}).

\hfill

It follows simply from (\ref{lie2}) y (\ref{rt}) that there 
exists a reduced
covariant derivative
\\[2 mm]
\be \label{rcd}
\triangledown v \equiv J(A)\D V
\ee
\\[2 mm]
corresponding to a {\it reduced connection} given by:
\\[2 mm]
\be
\label{rc} 
\gamma\ic=J(-A)^{\alpha}_{\mu} J(-A)^{\beta}_{\nu}
J(A)^{\rho}_{\delta}(\Gamma^{\delta}_{\alpha\beta} -
\d_{\alpha}J(A)^{\delta}_{\beta})
\ee
\\[2 mm]

With the definitions giving above, (\ref{rt}) and (\ref{rcd}),
the operation that gives the reduced tensors commutes with the 
basic operations of the tensor calculus: linear combinations, tensor 
products, contractions, permutation of indices and covariant 
derivation. That feature together with its simplicity (see \cite{b} )
is the reason to call the whole setting  {\it the Reduced Geometry}.

\hfill

In a Riemannian manifold with abelian Killing vectors

\begin{displaymath}
G_{\mu\nu} = \left( \begin{array}{cc}
G_{ab} & A_{ai} \\
A_{bj} & \,\,\,\,\,\,\hat{G}_{ij} + A_{ic}A_{jd}G^{cd} \\
 \end{array}  
\right) 
\end{displaymath}
with
\be
\label{lg} 
{\cal L}_{k_a}G_{\mu\nu}=0
\ee
there is a natural gauge field (\ref{gauge}) namely,
\be
\label{tgf}
A^{a}_{i}(x^j) = G^{ab}A_{bi}(x^j)
\ee

where $G_{ab}G^{bc}=\delta_{a}^{c}$.

\hfill

The {\it reduced Levi-Civita connection}, $\gamma_{l-c}$ has a 
non-zero torsion which
is the responsible for the {\it reduced curvature}  not being simply the
curvature 
of the reduced connection, as can be seen  in \cite{b}. 
In the general case, $\Gamma = 
\Gamma_{lc} + H$, the resulting reduced curvature is 

\be
\label{rc4}
r^{\eta}_{\lambda\delta\pi}= R(\gamma_{l-c}+h)^{\eta}_{\lambda\delta\pi} 
-2T(\gamma_{l-c})^{a}_{\lambda\delta} (\gamma_{l-c}+ h)^{\eta}_{a\pi}
\ee

where $h_{\mu\nu}^{\rho}$ is the reduced tensor corresponding to 
$H_{\mu\nu}^{\rho}$ and $T(\gamma_{l-c})$ is the Levi-Civita 
{\it reduced} torsion.

\subsection{Buscher's formulas for n commuting Killings}

The context of most applications of the formalism 
starts from a two-dimensional non-linear sigma model with target space
$\cM$,
whose bosonic part is given by:
\be
S=\int(G_{\mu \nu} +B_{\mu \nu})\d_{+}X^{\mu} \d_{-}X^{\nu}
\ee

The generalization of  Buscher's transformations 
to the case where there are
n commuting Killings present 
(${\cal L}_{k_a}G=0$ and ${\cal L}_{k_a}B=dW$)
follows easily from the {\it gauging procedure} \cite{aagl} .
The resulting dual model has the following
backgrounds :

\bea
\label{BF}
&\t Q^{\pm}_{ab} = Q^{ab}_{\pm} \nonumber\\
&\t Q^{\pm}_{ai} = \pm Q^{ab}_{\pm}Q^{\pm}_{bi} \nonumber\\
&\t Q_{ij} = Q_{ij} - Q^{cd}_{+}Q^{-}_{ci}Q^{+}_{dj}
\eea

with the conventions
$Q^{\pm}_{ab} \equiv G_{ab} \pm B_{ab}$,
$Q_{ab}^{\pm}Q^{bc}_{\pm}=\delta_{a}^{c}$,
$Q^{\pm}_{ai} \equiv G_{ai} \pm B_{ai}$ and 
$Q_{ij} \equiv G_{ij} + B_{ij}$ and  $Q^{ab}_{\pm}Q^{\pm}_{bc} =
\delta^{a}_{c}$. \\

The {\it quotient metric} is invariant under T-duality:
\be
\t {\hat{G}}_{ij} =  \hat{G}_{ij}
\ee
The three form $H \equiv d B$ is defined in tensorial terms as
\be
\label{H}
H_{\mu\nu\rho}\equiv {1\over2}(\D_{\mu}B_{\nu\rho}+
\D_{\nu}B_{\rho\mu}+
\D_{\rho}B_{\mu\nu})
\ee

where the Levi-Civita connection is used to define covariant derivatives.
Then, from (\ref{rt}) and (\ref{rcd}), 
the explicit computation yields for its {\it reduced} 
partner 
\bea
\label{3formredu}   
&h_{abc} = 0 ;\,\,\,\,\,\,\,h_{iab} = \frac{1}{2}\d_{i}B_{ab} ; 
\nonumber\\
&h_{aij} = -\frac{1}{2} F_{ij}^{a}(B)-\frac{A^{b}_{i}}{2}\d_{j}B_{ab} +
\frac{A^{b}_{j}}{2}\d_{i}B_{ab}; \nonumber\\
&h_{ijk} = \hat{h}_{ijk}
\eea
where $F^{a}_{ij}(B) \equiv \d_{i}B_{aj} - \d_{j}B_{ai}$.
The reduced three-form $h_{ijk}$ is actually T-invariant .

\section{The Canonical Map}

The classical string dynamics is goberned by the pullback of the
generalized
connection with torsion $\Gamma^{\pm}=\Gamma_{lc}\pm H$,
and at one-loop level the beta-functions of the bosonic and $N=1$ 
supersymmetric string models are proportional to the Ricci tensor
of that generalized connection.
Therefore our fist interest is to determine
the T-dual of the connection  $\Gamma^{\pm}$ (denoted as usual by
$\gamma_{\pm}$ when reduced) in the {\it reduced} setting:

\bea
\label{gct}
&\Gamma^{\pm\rho}_{\mu\nu} =& \Gamma^{\rho}_{(lc)\mu \nu} \pm
H_{\mu\nu}^{\rho}\\
&\gamma^{\pm\rho}_{\mu\nu} =& \gamma^{\rho}_{(lc)\mu \nu} \pm
h_{\mu\nu}^{\rho}
\eea

where the torsion $H_{\mu\nu\sigma}=
H_{\mu\nu}^{\rho}G_{\rho\sigma}$ is given by (\ref{H}). 
\\
To be specific, the starting point is:

\bea
&\g^{\pm \,\,c}_{ab} = 0 \nonumber\\
&\g^{\pm \,\,i}_{ab} =
 -{1\over2}\hat{\d}^{i}Q^{\mp}_{ab} \nonumber\\
&\g^{\pm \,b}_{ia} = {1\over2}G^{bc} \d_{i} Q^{\pm}_{ac}
=\g^{\mp \,b}_{ai}  \nonumber\\
&\g^{\pm \,j}_{ai} = {1\over2} \hat{G}^{jk} C^{\pm}_{ika}
=\g^{\mp \,j}_{ia}  \nonumber\\
&\g^{\pm \,a}_{ij} = -{1\over2}G^{ab}C^{\pm}_{ijb} \nonumber\\
&\g^{\pm \,\,k}_{ij} = \hat{\Gamma}^{\pm \,k}_{ij} \pm
\hat{h}^{\,\,\,k}_{ij} \equiv \hat{\Gamma}^{k}_{ij}
\eea
\\[2 mm]
where $C^{\pm}_{ijb} =
F_{ijb}(Q^{\pm}) + A^{d}_{i}\d_{j}Q^{\pm}_{bd} -
A^{d}_{j}\d_{i}Q^{\pm}_{bd}$.

\hfill

The T-duals are \footnote{It is exceedingly convenient
to take advantage of the transformation properties of the combination
$s_{ija}=\d_{i}Q^{\pm}_{ja}-(\d_{i}Q^{\pm}_{ab})A^{b}_{j}$
namely, $\t {s}_{ij} = \pm \frac{1}{Q_{\pm}}s_{ij}$ with $C_{ija}=
s_{[ij]a}$}:  

\be
\label{tdc2}
\t {\g}^{\pm\,\,\sigma}_{\mu \nu} =
t^{\mp\,\alpha}_{\mu}t^{\pm\,\beta}_{\nu}t^{\sigma}_{\pm\,\lambda}
\g^{\pm\,\,\lambda}_{\alpha \beta}
\ee

\hfill

for all components except $\t{\g}^{\pm\,\,b}_{ia}=\t{\g}^{\mp\,\,b}_{ai}$,
with

\bea
&t^{\pm\,b}_{a}=\pm Q^{ab}_{\pm};
&t^{\pm\,j}_{i} = t^{j}_{\pm i} = \delta^{j}_{i} \nonumber\\
&t_{\pm\,b}^{a}=\pm Q^{\mp}_{ab} \,\,\,& others = 0
\eea

The simplicity of the {\it reduced } transformations 
(\ref{tdc2}), allow us to built a map between the original ($\Sigma$)
and dual ($\t{\Sigma}$) geometries, {\it the canonical map}, 
transforming tensors ($V\rightarrow \t{V}$)
with the property of  mapping the corresponding
covariant derivatives linearly ($\t{\D}^{\pm}\t{V}\propto \D^{\pm}V$) :

\be
\label{canoni}
\t V^{\pm\ict}_{\icv} =
(\prod_{r=1}^{l} T^{\mu_r}_{\pm\,\beta_r}) 
(\prod_{s=1}^{m} T^{\pm\,\alpha_s}_{\nu_s})
V^{\pm \beta_1 ,...,\beta_l }_{\alpha_1 ,...,\alpha_m }
\ee

\hfill

where the $T^{\pm}$ and $T_{\pm}$ can be viewed as a sort of
vierbeins relating indices of the initial and dual geometries.
The covariant derivatives map lineraly but {\it not canonically},

\be
\label{casicanonide}
\t\D_{\rho}^{\pm}\t V^{\pm\ict}_{\icv} =
T^{\mp\,\lambda}_{\rho}(\prod_{r=1}^{l}T^{\mu_r}_{\pm\,\beta_r}) 
(\prod_{s=1}^{m}T^{\pm\,\alpha_s}_{\nu_s})
\D_{\lambda}^{\pm}V^{\beta_1 ,...,\beta_l }_{\alpha_1
,...,\alpha_m }
 \ee

\hfill

because the {\it anomaly} in the derivative index.
The matrices $T_{\pm}$ and $T^{\pm}$, 
first used by Hassan  \cite{hassan}, are :

\[ T^{\pm\,\nu}_{\mu} = \left( \begin{array}{cc}
\pm {Q^{ab}_{\pm}} & 0 \\
-Q^{ab}_{\mp} Q^{\mp}_{bi}\,\,\,\, & \delta^{i}_{j}
\end{array}
\right) \]
   
\[ T^{\mu}_{\pm \, \nu} = \left ( \begin{array}{cc}
\pm {Q^{\mp}_{ab}} & \pm Q^{\mp}_{ai} \\
0 \,\,\,\, & \delta^{i}_{j}
\end{array}
\right ) \]
\\[2 mm]

\hfill

$\nu$ being the column index and  $\mu$ the row index.
The above mentioned {\it anomaly} in fact implies that the
covariant derivative does not commute with the {\it canonical map},
and as a simple corrollary the curvatures do not transform simply by 
trading indices of  $T^{\pm}$ as we shall see next.

\hfill

Covariantly constant tensors (with respect to $\D_{\mu}^{\pm}$)
transform neccessarily as (\ref{canoni}). This is the case for the 
metric ($\D^{\pm}_{\mu}G = \t{\D}^{\pm}_{\mu} \t{G} = 0$),
for the holomorphic complex structures underlying extended 
supersymmetries \cite{hassan}, p-forms associated to W-algebras
\cite{kim}, and whatever holomorphic covariantly constant tensor we
found in our geometry.

\hfill
 
There are other tensors with non-canonical transformations, such as
the 2-form
$B_{\mu \nu}$ and the 3-form $H_{\alpha \beta \gamma}$.
Torsion is best studied as forming part of the generalized connection.
Actually, both equations  (\ref{canoni}) and (\ref{casicanonide}) together 
easily yield:
\\[1 mm]
\be
\label{conec1}
\t \Gamma^{\pm\,\rho}_{\mu\nu} = T^{\mp\,\lambda}_{\mu}
T^{\pm\,\beta}_{\nu}
T_{\pm\,\alpha}^{\rho} \Gamma^{\pm\,\alpha}_{\lambda\beta} +
(\d_{\mu} T^{\pm\,\beta}_{\nu}) T_{\pm\,\beta}^{\rho}
\ee
\\[1 mm]

The target-space connection transforms as a real T-duality connection
except for the anomaly in the $\mu$-index.

\newpage

\section{ Transformation of the Curvature and Canonical Connection}

Let us now consider the generalized curvature,  $R^{\pm}_{\mu\nu\sigma\rho}
= R(\Gamma^{\pm})^{\lambda}_{\mu\nu\sigma} G_{\lambda\rho}$
which obviously satisfies the symmetry relationships
$ R^{\pm}_{\mu\nu\sigma\rho}= -
R^{\pm}_{\nu\mu\sigma\rho}
= - R^{\pm}_{\mu\nu\rho\sigma}$, $ R^{\pm}_{\mu\nu\rho\sigma}=
R^{\mp}_{\rho\sigma \mu\nu}$ .
Working with the transformations of the connection in (\ref{tdc2}) 
and using the expression for the {\it reduced curvature} (\ref{rc4})
we get \footnote{$k^{a}_{\mu}\equiv k_{a}^{\nu}G_{\nu\mu}$.} :

\be
\label{curin}
\t {R}^{\pm}_{\mu \nu \sigma \rho} = T^{\mp \alpha}_{\mu} T^{\mp  
\beta}_{\nu} T^{\pm \lambda}_{\sigma} T^{\pm \delta}_{\rho} (
R^{\pm}_{\alpha \beta \lambda \delta} - 2 Q^{ab}_{\mp} \D^{\pm}_{\alpha}
k^{a}_{\beta} \D^{\mp}_{\lambda} k^{b}_{\delta} )
\ee  
\\
The fact that there is an inhomogeneous part ( $- 2 Q^{ab}_{\mp}
\D^{\pm}_{\alpha}
k^{a}_{\beta} \D^{\mp}_{\lambda} k^{b}_{\delta}$) in the transformation
of the curvature turns out to be a useful clue.\\
Actually, when an object does transform inhomogeneously, such as:
$\t{r} = \pm\prod t^{\Delta}( r + \psi)$, \footnote{We denote as $\prod
t^{\Delta}$ the product of matrices $t$ with $\Delta$  $\pm$ or
$\mp$ as appropiate} the involutive property, $T^2 = 1$, completely
determines
the transformation of the inhomogeneous part, $\psi$ ,namely  $\t{\psi} =
\mp \prod
t^{\Delta} \psi$. 
This fact inmediatly suggests the definition $w \equiv r+{1 \over 2} \psi$
, which does transform homogeneously.\\
There is then a quantity, the {\it corrected generalized curvature}
$W^{\pm}_{\alpha\beta\delta\eta}$ which transforms linearly:
\bea
\label{curcorre}
&W^{\pm}_{\mu\nu\sigma\rho}\equiv
R^{\pm}_{\mu\nu\sigma\rho} -
Q^{ab}_{\mp} \D^{\pm}_{\mu}
k^{a}_{\nu} \D^{\mp}_{\sigma} k^{b}_{\rho}\\
&\t{W}^{\pm}_{\mu\nu\sigma\rho} = T^{\mp \alpha}_{\mu} T^{\mp
\beta}_{\nu} T^{\pm \lambda}_{\sigma} T^{\pm \delta}_{\rho}
W^{\pm}_{\mu\nu\sigma\rho}
\eea
This correction  (\ref{curcorre}) is minimal because it includes
first-derivative terms and therefore it cannot be absorbed 
in a background's redefinition.

We have just seen that the transformations of the ordinary connection
is not canonical (\ref{canoni}). It is possible, however, to define a {\it
new}
connection with canonical transformation properties (\ref{conec1}) 
We shall refer to it as {\it canonical connection}:
\bea
\label{rccbla}
&\bar{\Gamma}^{\pm \,\,\rho}_{\mu \nu} = \Gamma^{\pm \,\,\rho}_{\mu \nu}
- G^{ab} k^{a}_{\mu} \D^{\mp}_{\nu} k^{b \, \rho}  \nonumber\\
&\t{\bar{\Gamma}}^{\pm\,\rho}_{\mu\nu} = T^{\pm\,\lambda}_{\mu}
T^{\pm\,\beta}_{\nu}
T_{\pm\,\alpha}^{\rho}
\bar{\Gamma}^{\pm\,\alpha}_{\lambda\beta} +
(\d_{\mu} T^{\pm\,\beta}_{\nu}) T_{\pm\,\beta}^{\rho}
\eea
%

Therefore the {\it canonical} covariant derivative, 
$\bar{\D}^{\pm}_{\mu}$ 
does now commute with the canonical mapping. This means that a
canonical transformation for tensors
\be
\label{tdcc12x}
\t V^{\pm\ict}_{\icv} =
(\prod_{A=1}^{m} T^{\pm\,\beta_A}_{\nu_A})
(\prod_{B=1}^{l} T^{\mu_B}_{\pm\,\alpha_B})
V^{\alpha_1 ,\ldots ,\alpha_l }_{\beta_1 ,\ldots ,\beta_m }
\ee

corresponds with a canonical one for the covariant derivatives :

\be
\label{blabla}
\t{\bar{\D}}_{\rho}^{\pm}\t V^{\pm\ict}_{\icv} =
T^{\pm\,\lambda}_{\rho}(\prod_{A=1}^{m}T^{\pm\,\beta_A}_{\nu_A})
(\prod_{B=1}^{l}T^{\mu_B}_{\pm\,\alpha_B})
\bar{\D}_{\lambda}^{\pm}V^{\alpha_1 ,\ldots ,\alpha_l }_{\beta_1
,\ldots \beta_m }
\ee

The {\it canonical} connection is compatible with the metric, 
$\bar{\D}^{\pm}_{\mu}G_{\nu\rho}=0$ provided that the Killing 
condition ${\cal L}_{k_a}G_{\mu\nu}=0$ is satisfied. 
Moreover, $\bar{\D}_{0}=0$ acting on $\Sigma$, impliying
$ k_{a}^{\alpha}\bar{R}^{\pm}_{\alpha \nu \sigma \rho} = 0$. 
At the end of this section we will give an intrinsic ( T-duality
independent) 
characterization of this {\it canonical connection} for which the above
properties will appear natural.

Also, following simply from the commutativity, we get the canonical
transformation 
of the curvature:
\be
\label{tdcc12}
\t{\bar{R}}^{\pm}_{\mu\nu\sigma\rho}= T^{\pm\,\alpha}_{\mu}
T^{\pm\,\beta}_{\nu}
T^{\pm\,\delta}_{\sigma} T^{\pm\,\eta}_{\rho}
\bar{R}^{\pm}_{\alpha\beta\delta\eta}
\ee
\\
With the {\it canonical connection}, the {\it canonical T-duality map} 
commutes with the
basic operations of the tensor calculus, i.e., linear combinations, tensor
products, 
permutation and contraction of indices and covariant derivation.
In particular it implies that every tensor built from 
${\bar{R}}^{\pm}_{\mu\nu\sigma\rho}$, $G_{\alpha\beta}$, 
$\bar{\D}^{\pm}_{\eta}$, and any other tensor transforming canonnically,
{\it transforms canonically}.
As a corollary, target-space canonical scalars are T-duality scalars (
$\t{\bar{R}}^{\pm}=\bar{R}^{\pm}$,
$(\t{\D}^{\pm}\t{R}^{\pm})^2 =({\D}^{\pm}{R}^{\pm})^2$,
$\t{\bar{R}}^{\pm}_{\mu\nu}\t{\bar{R}}^{\pm\,\,\mu\nu}=
\bar{R}^{\pm}_{\mu\nu}\bar{R}^{\pm\,\,\mu\nu}$, 
$\t{\bar{R}}^{\pm}_{\mu\nu\sigma\rho}\t{\bar{R}}^{\pm\,\,\mu\nu\sigma\rho}=
\bar{R}^{\pm}_{\mu\nu\sigma\rho}\bar{R}^{\pm\,\,\mu\nu\sigma\rho}$ ...).

\hfill

In {\it complex manifolds}, the presence of additional {\it canonical
tensors},
i.e., the holomorphic complex structures $J^{\pm\,\,\mu}_{\nu}$,
allows the construction of another
{\it new} set of T-duality scalars
($\bar{R}^{\pm}_{\mu\nu}J^{\mu\nu}_{\pm}$,
$\bar{R}^{\pm}_{\mu\nu\sigma\rho}J^{\mu\sigma}_{\pm}J^{\nu\rho}_{\pm}$,
$\bar{R}^{\pm}_{\mu\nu\sigma\rho}J^{\rho\alpha}_{\pm}\bar{R}_{\alpha}^{
\pm\,\,\mu\nu \sigma}$,
...).

\hfill

The definition of {\it canonical connection} in (\ref{rccbla}) was
motivated
in its very convenient transformation properties under T-duality.
Nevertheless,
an intrinsic (T-duality independent ) characterization can be given for it.

\hfill

Let us start with our set of abelian (Killing) vectors 
$\{k^{\mu}_{(a)}\}$\footnote{$k_{(a)}^{\mu}\equiv k_{a}^{\mu}$.}
and the vector space {\cal K}
spanned by them. The presence of a metric 
$G_{\mu\nu}$ on our manifold 
induces the natural projector on {\cal K}, 
$P_{\mu}^{\nu}\equiv k_{(a)\mu}k^{\nu}_{(b)}G^{ab}$
\footnote{We remember $G_{ab}= k^{\nu}_{(b)}k_{(a)}^{\nu}$ and
$G^{ab}G_{bc}=\delta_{c}^{a}$}, with 
$P^2=P$ and $Pk_{(a)}= k_{(a)}$.

\hfill

Now, let us  take the quotient of the whole space 
of connections {\cal C} by the projection $P$,i.e., ${\cal C}/P$.
Then, two connections $\Gamma^1$ and $\Gamma^2$ belong to the
same class on ${\cal C}/P$ if $(\Gamma^{1}-\Gamma^{2})_{\mu}=
P_{\mu}^{\nu}L_{\nu}$ for some matrix valued one-form $L$.

\hfill

In every class there is an unique covariant derivation ,$\bar{\D}$,
with the property 

\be
\bar{\D}_{k_{(a)}}=
{\cal L}_{k_{(a)}}
\ee

\hfill  

\footnote{$\bar{\D}_{k_{(a)}}\equiv k^{\mu}_{(a)}\bar{\D}_{\mu}$}
Writing $\bar{\D}=P^{\bot}\bar{\D}+ P\bar{\D}$, the ortogonal component
($P^{\bot}=1-P$) is common to every element on the same class, 
say ${\D}^{\bot}$, and the {\cal K} projection is
$(P\bar{\D})_{\mu}= k_{(a)\mu}k^{\nu}_{(b)}G^{ab}\bar{\D}_{\nu}=
k_{(a)\mu}G^{ab}{\cal L}_{k_{(b)}}$. 
Therefore, the {\it barred connection} is

\be
\label{barredbla}
\bar{\D}_{\mu} ={\D}^{\bot}_{\mu} + k_{(a)\mu}G^{ab}{\cal L}_{k_{(b)}}
\ee

in every class of ${\cal C}/P$.

If we start in a class having a connection compatible with the metric (as
it
is the case in T-duality (\ref{gct})), the compatibility of the {\it barred
connection}
trivially implies the Killing condition ${\cal L}_{k_{(a)}}G_{\mu\nu}=0$.

\hfill

In adapted coordinates, the $\bar{\D}_{k_{(a)}}={\cal L}_{k_{(a)}}$
condition
means $\bar{\D}_{a}=\partial_{a}$ and then
$\bar{\Gamma}_{a\,\nu}^{\rho}=0$.
Taking an arbitrary reference connection in the class, $\Gamma$ ($\D$)
the above conditions imply

\be
\label{bla123}
\bar{\Gamma}_{\mu\nu}^{\rho}=\Gamma_{\mu\nu}^{\rho}-
k_{(a)\mu}G^{ab}(\D)^{\aleph}_{\nu}k_{(b)}^{\rho}
\ee

where the $\aleph$ {\it operation }on connections simply flips the sign
of the torsion, 
$\aleph : \Gamma^{\rho}_{(\mu\nu)}\rightarrow \Gamma^{\rho}_{(\mu\nu)}$
and
$\aleph : \Gamma^{\rho}_{[\mu\nu]}\rightarrow -\Gamma^{\rho}_{[\mu\nu]}$.
Note that the $\aleph$ {\it operation} transforms our stringy connections 
$\Gamma^{\pm}$ one into the other. In that way, {\it the barred connection}
(\ref{barredbla}) (\ref{bla123})
agree with the {\it T-duality canonical} one (\ref{rccbla}) 
if we  choose the classes which $\Gamma^{\pm}$belong to.

\hfill

With respect to the reference connection $\Gamma$,
the {\it barred connection} loses the information about 
the parallel transport in the Killing's direction. This {\it ortogonal
projection} is easily seen rewriting (\ref{bla123})
in a non-covariant form 

\be
\label{opbla}
\bar{\Gamma}_{\mu\nu}^{\rho}=
P^{\bot\,\alpha}_{\mu}\Gamma_{\alpha\nu}^{\rho}-
k_{(a)\mu}G^{ab}\partial_{\nu}k_{(b)}^{\rho}
\ee

As a consequence of  the defining properties, 
$k_{(a)}^{\alpha}\bar{\D}_{\alpha}k_{(b)}^{\mu}
={\cal L}_{k_{(a)}}k_{(b)}^{\mu}=0$, implying
the {\it barred geodesics}

\be
\label{bgblabla}
\ddot{X}^{\mu}+ 
\bar{\Gamma}_{\alpha\beta}^{\mu}\dot{X}^{\alpha}\dot{X}^{\beta}
=0
\ee

have 
{\it the free motion on the Killing's direction}

\be
\label{fmkdbla}
\dot{X}^{\mu}=C^{a}k_{(a)}^{\mu}(X(\tau))\,\,\,\,\,;
\,\,\,\dot{C}^{a}=0
\ee

as a consistent solution. Let us note this motion is allowed
for {\it every} barred connection provided that it projects out
the parallel transport in the {\cal K} direction.

\hfill

Finally let us remark that $(\bar{\D})^{\aleph}k=0$. Of course,
the $(\bar{\D})^{\aleph}$ is in general non-compatible with
the metric. As a corollary, we get the
useful (see next section) condition 

\be
\label{usefulbla}
k_{(a)}^{\alpha}\bar{R}_{\alpha\mu\nu}^{\sigma}=
({\cal L}_{k_{(a)}}\bar{\D})_{\mu\nu}^{\sigma}
\ee

\hfill

which for the stringy connections means 
\be
\bar{R}^{\pm}_{a\mu\nu\rho}=0
\ee

\section{Invariant Polynomials}

Invariant polynomials $P(\Omega)$ are characterized in general (cf
\cite{egh})
by $P(\Omega) = P(g^{-1}\Omega)g)$, (where   $\Omega^{\rho}_{\sigma} 
\equiv R(\Gamma)^{\,\,\,\rho}_{\mu \nu
\sigma} d x^{\mu} \wedge d x^{\nu} $ 
is the matrix-valued 
curvature two-form), which implies that $d P(\Omega) = 0$; and moreover,
that 
$P(\Omega)$ has topologically invariant integrals (on manifolds without
boundary). Chern and Simons
have proven the specific result that given two different connections,
$\omega$
and $\omega '$
\be
P(\Omega ') - P(\Omega) = d Q(\omega ',\omega)
\ee
where $Q$, the Chern-Simons term, is given by $Q(\omega ',\omega) \equiv
r \int_{0}^{1} P(\omega ' - \omega,\Omega_t,\ldots ,\Omega_t) $, r being
the degree of the 
polynomial, and $\Omega_t \equiv d \omega_t + \omega_t \wedge \omega_t $,
with
$\omega_t \equiv t \omega ' + (1-t) \omega$.
\\
As a consequence, all those polynomials can be determined (up to a total
differential,
that is in a cohomological sense), using {\it any} convenient connection;
in our case
the {\it canonical connection} ${\bar\Gamma}^{\pm}$ 
imposes itself naturally, because as
 we have seen in detail, the corresponding curvature, 
${\bar R}^{\pm \,\,\rho}_{\mu \nu\sigma} $ transforms canonically under
T-duality.
 
It is plain that the only non-zero components of any {\it canonical
invariant
polynomial} would be the ones with all 
indices transverse. This is clear ,because ${\bar\nabla}_{a}^{\pm} = 0$
actually
implies ${\bar R}_{a\alpha\beta}^{\delta} = 0$.\\

In the particular case in which we are considering a Pontryagin
characteristic
polynomial, appropiate when the curvature lies in the Lie algebra of
$O(k)$;
$p_j (\Omega)\in \Lambda^{4j}(M)$ we can write 
$k^{a) \,\mu} \bar{P}^{(4j)}_{\mu \nu_{1} ...\nu_{4j-1}} = 0 \,\,\,\,  
\forall a,\nu_{1}...\nu_{4j-1}$ or, in adapted coordinates,
\be
\label{top}
\bar{P}^{(4j)}_{a \nu_{1}...\nu_{4j-1}} = 0
\ee
Now, $\bar{P}$ transforms canonically:
\be
\t{\bar{P}}^{(4j)}_{\nu_{1} ...\nu_{4j}} = (\prod_{A=1}^{4j} T^{\pm
\beta_{A}}_{\mu_{A}}) \bar{P}^{(4j)}_{\beta_{1}...\beta_{4j}} 
\ee

And for the non-vanishing components:

\be
\label{bbla}
\t{\bar{P}}^{(4j)}_{i_{1} ...i_{4j}} = \prod_{A=1}^{4j}
\delta^{k_{A}}_{i_{A}}
\bar{P}^{(4j)}_{k_{1}...k_{4j}} = \bar{P}^{(4j)}_{i_{1}...i_{4j}}
\ee

This means that the components of the {\it canonical} Pontryagin forms
are actually invariant.
A glance at (\ref{top}) implies that Top Forms vanish
(because they necessarily include Killing indices, for which
$\bar{R}^{\pm} = 0) $ . 
\\
This in turn means that if the canonical connection has a global meaning
 ( $det(G^{ab} \neq 0)$ the topological invariants obtained by integrating
top forms are necessarily zero, both in the original and in the dual model.
\\
Chern classes are defined for complex manifolds with $\Omega \in
gl(k,\mathbb{C})$;
$c_j\in \Lambda^{2j}(M)$, but are otherwise similar to the Pontryagin ones
from
the point of view of the present work.
\\
A well known fact is that for even dimensional manifolds a further
$SO(2r)$ invariant polynomial can be defined (the Pfaffian). The
corresponding
 Euler class is essentially the square root of the highest Pontryagin
class.
The mother of all index theorems is precisely the Gauss-Bonnet theorem,
which states that
the  Euler characteristic $\chi(M)$ is  integral of the Euler
class $e(\bar{R}^{\pm})$ . Now we see that the integrand (a top form) is 
necessarily cero if the group G acts
freely (without fixed points). From (\ref{bbla}), this assertion 
is T-duality invariant
\\
When there are fixed points, we could define a one parameter family of
connections compatible with the metric \footnote{Because of the Killing 
condition.} interpolating from
 $\bar{\Gamma}^{\pm}$ at $t=1$ to
${\Gamma}^{\pm}$ en $t=0$.
\\
\be
\Gamma^{\pm \sigma}_{\mu \nu}(t) = \Gamma^{\pm \sigma}_{\mu \nu} - \frac{t
k_{\mu}}{k^2 + (1-t)^2} \D^{\mp}_{\nu} k^{\sigma}
\ee
\\
This is well defined $\forall t>0$
whereas in the limit  $t \rightarrow 0$ the top forms 
vanish everywhere except perhaps at the fixed points.
This means that all the topological information is stored in the fixed
points.
\\
In the case at hand this is contained in the well-known theorem by
"Poincar\'e-Hopf"  asserting that in a compact manifold
$M$ endowed with a differentiable vector field, $w$ (Killings in our case)
with isolated zeroes, the sum of the corresponding indices $\iota$ (that is,
the Brouwer degree of the mapping $\hat{k}(x)\equiv \frac{k(x)}{||k(x)||})$
\cite{milnor} is precisely 
Euler's characteristic:

\be
\chi(M)\,=\,\sum \iota = \,\sum_{i=0}^{m}(-1)^i \,\,\,\,rank H_i(m).
\ee
(where $H_i(m)$ stands for the  i-th homology group of $M$.)
\\
As a trivial corollary, only in manifolds with zero Euler Characteristic it
is
possible to have free Killing actions.\\

Owing to the fact that we do not have enough control on the topology of the
dual manifold, we can not say anything about the topological numbers, 
which are {\it integrals } of the invariant polynomials, except, of course,
when
they vanish. It would also be quite complicated to keep track of all
boundary terms
in manifolds with non-trivial boundary in order to compute, say, the $\eta$
invariant.
\par
\par

{\bf Acknowledgements.}
We are grateful to  Luis Alvarez-C\'onsul, Luis Alvarez-Gaum\'e, 
 C\'esar G\'omez, Tom\'as Ort\'in
and
Kostas Sfetsos .
This work was supported by
the UAM (JHL), CAM(JB) ; CICYT AEN/96/-1655 and AEN/96/1664
and an EU TMR contract ERBFMRXCT960012.

\end{document}